\documentclass[12pt]{article}
\usepackage{graphicx}
\usepackage[top=2cm, bottom=3cm, left=3cm, right=3cm]{geometry}

\title{The Local Bubble in the interstellar medium and the origin of the low energy cosmic rays}
\author{$^{*}$A.D. Erlykin $^{1,2}$, S.K. Machavariani $^{1}$, A.W. Wolfendale $^{2}$ \\
$(1)$ P N Lebedev Physical Institute, Moscow 119991, Russia \\
$(2)$ Physics Department, Durham University, Durham, DH1 3LE, UK }

\begin{document}
\maketitle

\footnote{$^{*}$Corresponding author: tel +74991358737 \\ 
 E-mail addresses: erlykin@sci.lebedev.ru; machavar@sci.lebedev.ru; a.w.wolfendale@durham.ac.uk}

\begin{abstract}
An analysis of the energy spectra of cosmic rays and particularly the precise data from the AMS-02
 experiment support the view about the important role of the Local Bubble in the nearby interstellar medium. It is 
suggested that the bulk of CR below about 200 GV of rigidity (momentum/charge ratio) comes from the modest number of 
supernova remnants in the Local Bubble which appear to have occurred some 10$^6$years ago and contributed to its 
formation. At higher rigidities the contribution from a 'Local Source', a single supernova remnant generated some 
10$^5$ years ago seems to dominate up to, at least 1000 GV.
\end{abstract}   

\vspace{1mm}

{\bf Keywords:} cosmic rays, local bubble, energy spectrum

\section{Introduction}
The origin of cosmic rays (CR) is still subject of some doubt, although supernova remnants (SNR) appear to play an
 important part, at least to several PV (Ginzburg and Syrovatskii, 1964 and many other publications). Recent 
precise measurements of the rigidity spectra of 
protons and helium nuclei, the positron fraction, the antiproton to proton ratio and preliminary data on
 the rigidity spectrum of lithium and boron to carbon ratio (Ting for AMS-02 coll., 2015)
all support the previous view (Erlykin and Wolfendale, 2013, 2015a,b; Tomassetti, 2015a,b; Kachelriess et al., 2015) 
that there is a large contribution to the intensity from a nearby, recent SNR above 200 GV, particularly for the 
secondary nuclei (Erlykin and Wolfendale, 2016a; Tomassetti, 2015a,b).

We turn now to rigidities below 200 GV where earlier we proposed that a 'New Component' dominates (Erlykin and 
Wolfendale, 2012). This idea has been put forward more recently in (Tomassetti, 2015c,d) for, at least, carbon and 
oxygen. Thus, if the above is correct, the idea of CR at Earth being largely derived from very many SNR in the Galaxy 
needs revision.

Here we put forward the idea that the origin scenario is more complicated, with significant features due to many,
say, tens of SNR  (analogous to multiple scattering of particles), a modest, say, several number of SNR, locally 
(analogous to plural scattering) and a single SNR (analogous to a single Rutherford scattering). Here, the analogy 
with the classification of scatterings is given simply as a visual illustration. In the present paper we deal with the
 'modest number of SNR, locally'.
This involves an examination of the local environment of the Solar System in the Local Bubble, a region 
in which low energy CR isotopes have been hypothesised to be accelerated (Binns et al., 2007, 2016). 

Experimental
study of the CR isotope composition may provide information on the matter content and diffusion properties of the 
local ISM, but it is not easy because of the low density of such isotopes. It is particularly true for radioactive 
isotopes, such as $^{10}Be$, $^{26}Al$, $^{60}Fe$ and others, the low density of which can be due not only to low 
density within the Local Bubble, but by their decay too. The problem has been discussed in many papers, for example
in Ptuskin and Soutoul (1998).   
\section{The Local Bubble}
There is no doubt that the Solar System currently lies in a Local Bubble of effective radius ~200pc (Frisch, 1997)
. The Local Bubble contains hot gas of low density (10$^6$K; 10$^{-2}$cm$^{-3}$) (Frisch, 1997; Berghofer and 
Breitschwerdt, 2002) and was probably caused by a succession of supernovae (about 10) starting some 10$^6$ years ago 
and by associated stellar winds from massive stars. Such a number of SNe can be compared with that expected
 on the basis of the usually assumed galactic frequency of 10$^{-2}$ year$^{-1}$, an effective galactic radius of 
15kpc and a disk thickness of 0.3kpc, leading to an expected number of 2 for a uniform SNR distribution in space and 
time. Thus, the actual number of local SNR is a little smaller than expected. 

It must be realised, however, that the actual volume in which the supernova formed would have been much smaller than 
the 'radius 200 pc'. This is because the relevant OB Associations have radii of only about 20pc (Mel'nik and Efremov,
1995) and the actual number of 'local SNR' could have exceeded the number expected by a large factor ($\approx 
100$). What is certain is that the Solar System is in an unusually hot interstellar medium over a period of some 
$10^6$ years where there has been 'recent' supernova activity.
\section{The expected cosmic rays from the supernovae in the Local Bubble}
In Erlykin and Wolfendale, (2015a) we evaluated the proton spectra expected from Geminga and Monogem SNR and 
compared them with our estimate of the 'observed' spectrum from a single source, using the AMS-02 results. The 
latter was derived from the measured proton spectrum minus the assumed 'background'. Since this 'background' was 
defined as that due to CR produced by many old and distant SNR contributing mostly in the low energy region we fitted 
its spectrum in this region by its power law and extrapolated it to energies higher than that of the kink at 300 GV 
(Erlykin and Wolfendale, 2015a, Figure 2). The calculations of the spectra from Geminga and Monogem SNR  
were made for the situations of alternatively normal and anomalous diffusion. The anomalous diffusion is more 
appropriate for the non-uniform ISM with bubbles and variety of densities and temperatures. We expect the diffusion 
to be anomalous for the conditions inside the Local Bubble and for some distance beyond (Erlykin et al., 2003).
 The result was a prediction lower than 'observed' by a factor 30 at 1000 GeV for Geminga and anomalous diffusion.
Three factors combine to account for the intensity of the New Component: \\
(a) The fact, referred to in section 2, that there has been a succession of supernovae locally over the past $10^6$ 
to $10^7$ years. \\
(b) The increased temperature which would have started after the first few supernovae and  would have caused a 
considerable increase in the CR injection efficiency (Erlykin et al., 2016a). \\ 
(c) The existence of a Bubble implies the trapping of particles and plasma within. The Cygnus Bubble is an example 
(eg. Ackermann et al., 2011) and the Fermi Bubble near the Galactic Centre (Ackermann et al., 2014) is another.
Indeed, the spectral shape of the excess from the Fermi Bubble is similar to what we expect from our own New 
Component. The fact that the Galactic Explosion at the Galactic Centre some 5My ago gives a Bubble that is still 
visible - with its sharp edges - is in the spirit of the situation for our Local Bubble. It should be added that 
Higdon et al. (1998) stressed the role of Cosmic Ray acceleration from SN ejecta in Superbubbles nearly 20 years ago.
 
Reiterating (c) the problem is in dissuading the new CR from leaving the local interstellar medium; under 
normal conditions the CR generated some few million years ago would have diffused of several kpc by now. In this
 respect the Local Bubble can be invoked as a trapping volume. 

The increased injection efficiency with increased interstellar medium temperature is important (point (b)). In our 
work (Erlykin et al., 2016b) we explained the remarkably small CR intensity gradient in the Galaxy in terms of the 
radial gradient of the interstellar medium temperature. The injection efficiency was shown to increase very 
rapidly with increasing temperature. The result of relevance here is that the efficiency in the Local Bubble should be
 very high, the later supernovae benefitting from the increased interstellar medium
temperature arising from the earlier ones. Extrapolating the plots of the fraction of ambient protons of energy 
sufficient to be injected for CR acceleration as a function of temperature yields efficiencies approaching 100\% 
for the very high Local Bubble temperature (Erlykin et al., 2016b, Figure 4).    

It is postulated that it is not the 'current' single source which is important at 1000 GV, that is responsible 
for the CR below 200 GV ('the New Component') but rather the other old SNR which generated the Local Bubble. The 
antiquity of the SNR means that their contributions to the CR intensity will be biased towards lower energies,
the higher energy particles having left the Local Bubble. Lack of knowledge of the supernovae in excess and their 
ages coupled with uncertainty in the actual trapping efficiency of the Local Bubble precludes accurate calculations 
of the CR spectra at present. 

The high rate of recent SNR will also mean that the 're-acceleration' process is relevant, see, for example,
Thoudam and H\"{o}randel, (2014). The weak shocks associated with old SNR will also give significant 
extra fluxes of CR, mainly below 200 GV due to their larger volume. These particles will form part of the new 
component.
\section{Conclusion}
With data on the characteristics of the Local Bubble and the local interstellar medium, and our position within it, 
it is likely that a significant fraction of CR of rigidity 200 GV, and below (our 'New Component') comes from the 
claimed excess of supernovae responsible for the Bubble, in the last a few $10^6$ years.   

{\bf Acknowledgements}

We are grateful to the Kohn Foundation for financial support.

\end{document}